\documentclass[twocolumn,preprintnumbers,prx]{revtex4}
\usepackage{graphicx}
\usepackage{dcolumn}
\usepackage{amssymb}
\usepackage{bm}

\begin{document}

\title{Phase Space Quantum Mechanics as a Landau Level Problem}
\author{Kun Yang}
\affiliation{National High Magnetic Field Laboratory and Department of Physics,
Florida State University, Tallahassee, Florida 32306, USA}

\begin{abstract}

We point out the connection between the problem of formulating quantum mechanics in phase space and projecting the motion of a quantum mechanical particle onto a particular Landau level. In particular, we show that lowest Landau level wave functions, which are widely used in studies of quantum Hall effect, are actually phase space wave functions in this context. We demonstrate the usefulness of this understanding by analyzing some simple problems, and propose other utilities.

\end{abstract}

\date{\today}

\maketitle

\section{Introduction}

Quantum mechanical motion of particles are described by complex wave functions, usually written in position basis (or equivalently, real space). Square of the absolute value of such wave functions represents the probability density of finding the particle at the corresponding position. Over the years attempts have been made to formulate quantum mechanics (QM) in phase space, and in particular describe the motion of particle using phase space wave functions, which are functions of position {\em and} momentum. Conceptually such a formulation may allow for a more direct connection with classical mechanics (CM), which is formulated in phase space.
At a more practical level the phase space wave function allows one to talk about the {\em joint} distribution function of position {\em and} momentum, which is indeed relevant in many measurement schemes. While the community has witnessed significant formal developments along this direction, one also gets the impression that phase space QM has yet to demonstrate its power in solving the Schr\"{o}dinger equation for specific cases, where the traditional real space (or position basis) formulation appears to be easier to use.

One of the purposes of this paper is to point out that in the field of quantum Hall effect (QHE), where (to very good approximations) the electron motion is confined to a given Landau level (LL), people have actually been using phase space QM\cite{JainKivelson} and in particular the phase space wave function with great success, perhaps without realizing it! The other (more important) purpose is to deepen our understanding the phase space QM by using the insights gained from studies of QHE, and to propose possible utilities of these insights, especially in understanding the connection between QM and CM.

In the remainder of this paper we start by reviewing the LL problem (largely following Ref. \onlinecite{Book}) and pointing out its equivalence to the formulation of phase space QM initiated in Ref. \onlinecite{Torres-Vega and Frederick}, in the limit of large LL spacing. The usefulness of this understanding is then demonstrated by using it to analyse both the ordinary and inverted harmonic oscillator problems. We end with a proposal of formulating QM in terms of the nodes of the phase space wave functions.

\section{Landau level projection and phase space quantum mechanics}
Consider the Hamiltonian (kinetic energy only for the moment) of an electron confined to the xy-plane and subject to a uniform perpendicular magnetic field:
\begin{equation}
T=\frac{1}{ 2m}(\vec{p}+\frac{e}{ c}\vec{A})^2 = \frac{1}{ 2m}\vec{\Pi}^2=\frac{1}{ 2m}(\Pi_x^2+\Pi_y^2),
\label{eq:hamiltonianintermsofmechanicalmomentum}
\end{equation}
where
\begin{equation}
\vec{\Pi}=\vec{p}+\frac{e}{ c}\vec{A}(\vec{r})
\end{equation}
is the 2D {\em mechanical} momentum of the electron. Unlike the canonical momentum $\vec{p}$, the different components of $\vec{\Pi}$ do {\em not} commute:
\begin{equation}
[\Pi_x, \Pi_y]=-\frac{i\hbar e}{ c}(\partial_x A_y - \partial_y A_x)= \frac{i\hbar eB}{ c} = \frac{i\hbar^2}{ \ell^2},
\end{equation}
where $\ell = \sqrt{\frac{\hbar c}{eB}}$ is the magnetic length, and for convenience we choose
$\vec{B} = -B\hat{z}$. We thus find the commutation relation between $\Pi_x$ and $\Pi_y$ is similar to that between $x$ and $p_x$; thus Eq.~(\ref{eq:hamiltonianintermsofmechanicalmomentum}) takes the form of a 1D harmonic oscillator Hamiltonian.
We therefore introduce
\begin{equation}
a=\frac{\ell}{\sqrt{2}\hbar}(\Pi_x+i\Pi_y), \hskip 0.5 cm a^\dagger=\frac{\ell}{\sqrt{2}\hbar}(\Pi_x-i\Pi_y),
\end{equation}
yielding $[a, a^\dagger]=1$ and
\begin{equation}
T=\hbar\omega_c(a^\dagger a+1/2).
\label{eq:hamiltonianintermsofmechanicalmomentumsmg}
\end{equation}
We thus immediately obtain the Landau level spectrum
\begin{equation}
E_n=\hbar\omega_c(n+1/2),
\end{equation}
{\em without} specifying the gauge (here $\omega_c = \frac{eB}{mc}$ is the cyclotron frequency).
How do we see that the Landau levels have (massive) degeneracy? To this end we introduce another set of conjugate variables (similar to $\Pi_x$ and $\Pi_y$), known as guiding center coordinates:
\begin{equation}
\vec{R}=(R_x, R_y)=\vec{r}-\ell^2(\hat{z}\times\vec{\Pi})/\hbar,
\label{eq:guidingcentercoordiates}
\end{equation}
from which it is easy to show
\begin{equation}
[R_x, R_y]=-i\ell^2.
\label{eq:guidingcentercommutator}
\end{equation}
We can thus introduce another set of harmonic oscillator operators
\begin{equation}
b=\frac{1}{\sqrt{2}\ell}(R_x-iR_y), \hskip 0.5 cm b^\dagger=\frac{1}{\sqrt{2}\ell}(R_x+i R_y),
\end{equation}
with $[b, b^\dagger]=1$. More importantly,
\begin{equation}
[R_i, \Pi_j]=[R_i, T]=[a, b]=[a, b^\dagger]=0,
\label{eq:guidingcentercommutatorwithothers}
\end{equation}
which means $R_i$ only has non-zero matrix elements between states {\em within the same Landau level}, and so do $b$ and $b^\dagger$. We thus have two independent harmonic oscillators, described by operators $a$ and $b$, whose (number operator) eigenstates span the Hilbert space of a 2D quantum particle.

In QHE one often takes the limit $\hbar\omega_c \rightarrow \infty$ (by either sending $B\rightarrow \infty$ or more often theoretically, $m\rightarrow 0$), as a result the motion of electron is confined to a given (often the lowest) LL. Consequently the harmonic oscillator $a$ is frozen (say to its ground state), and the remaining degree of freedom is that of the harmonic oscillator $b$, which describes a 1D Hilbert space! To see this explicitly, we note
$\vec R$ is in fact the position operator $\vec{r} \equiv (x,y)$ {\em projected onto a given Landau level}:
\begin{equation}
\vec{R}=P_n\vec{r}P_n,
\label{eq:projected position operator}
\end{equation}
where $P_n$ is the projection operator onto the $n$th Landau level.
The degenerate states within the same Landau level can be generated using $\vec{R}$ or $b$ and $b^\dagger$. Again everything we said above is independent of the choice of gauge.
We thus find $x$ and $y$, which commute with each other and (along with their conjugate momenta $p_x$ and $p_y$) span the 2D Hilbert space, becoming a conjugate pair [cf. Eq. (\ref{eq:guidingcentercommutator})] like $x$ and $p_x$ themselves! In particular all LL projected operators can be expressed in terms of $R_x$ and $R_y$, just like all operators in 1D QM can be expressed in terms of $x$ and $p_x$. Thus once projection to a given LL is performed we have reduced a 2D QM problem to a 1D QM problem.

We are now in a position to discuss the relation between the LL problem discussed above with one particular approach to phase space formulation of QM\cite{Torres-Vega and Frederick,note1}. In order to construct phase space wave functions, Torres-Vega and Frederick\cite{Torres-Vega and Frederick} postulated the existence of an abstract Hilbert space spanned by a complete and orthonormal basis set $\{|\Gamma\rangle\} = \{|q, p\rangle\}$, which are simultaneous eigenstates of an abstract conjugate pair $(\hat{q}, \hat{p})$, which {\em commute} with each other in this abstract Hilbert space. The {\em physical} conjugate pair $(\hat{Q}, \hat{P})$, which are operators defined in the {\em physical} Hilbert space, have the non-trivial commutation relation $[\hat{Q}, \hat{P}] = i\hbar$, and can be written as proper combinations of $q, p$ and $\partial_q, \partial_p$.
Later on it was pointed out\cite{Ban} that the abstract Hilbert space can be constructed by supplement the physical Hilbert space by introducing an auxiliary (or ``relative") conjugate pair.
It should be clear by now that in the LL problem we just discussed, the pair $(\hat{q}, \hat{p})$ corresponds to $(x, y)$ which indeed commute with each other in the full Hilbert space, and $(\hat{Q}, \hat{P})$ corresponds to $(R_x, R_y)$ which are projected version of $(x, y)$ which no longer commute in an LL subspace. The physical Hilbert space (of a specific LL) is that spanned by the pair $(b, b^\dagger)$, while the auxiliary (or ``relative") conjugate pair correspond to $(a, a^\dagger)$. We thus find that QM in a LL, if viewed as QM in the original full Hilbert space, provides a physical realization of the phase space QM envisioned by Torres-Vega and Frederick.

\section{ Lowest Landau level wave function as phase space wave function}
As it should be clear by now, following the traditional QM approach a state in a specific LL should be described by a 1D wave function, written say in the $R_x$ basis. However in QHE one almost always works with 2D (real space) wave functions in the original 2D Hilbert space, even though they are actually constrained to a much smaller subspace, say the lowest LL (LLL). Obviously such wave functions must be heavily constrained, and such constraint is well-understood in QHE\cite{Book}: In the symmetric gauge, all LLL wave functions must take the form
\begin{equation}
\Psi(x,y) = f(z) e^{-\frac{1}{4}|z|^{2}}
\label{eq:analyticallowest Landau levelwavefunction}
\end{equation}
where $z=(x + iy)/\ell$ and $f(z)$ is an {\em analytic} function of $z$. The analytic structure of Eq. (\ref{eq:analyticallowest Landau levelwavefunction}) has been extremely helpful to understand various aspects of QHE, and we will suggest its applications in phase space QM later on. For the moment though we would like to point out that in the context of the present discussion, Eq. (\ref{eq:analyticallowest Landau levelwavefunction}) is actually a {\em phase space} wave function\cite{Harriman}, written as a function of $x$ and $y$, which have become a conjugate pair $R_x$ and $R_y$ after projecting to the LLL. In particular $|\Psi(x,y)|^2$ can be understood either as the probability density in the original 2D real space, or (in the present context) the {\em phase} space $(R_x, R_y)$ of the relevant 1D system! Note the uncertainty dictated by Eq. (\ref{eq:guidingcentercommutator}) is {\em not} violated here, due to the analytic constraint Eq. (\ref{eq:analyticallowest Landau levelwavefunction}) must satisfy\cite{Book}.


\section{Harmonic Oscillator Example}
In 1D QM one has a Hamiltonian of the form
\begin{equation}
H= \frac{p^2}{2} + v(x),
\label{eq:Original 1D Ham}
\end{equation}
which is equivalent to
\begin{equation}
H= \frac{R_y^2}{2} + v(R_x),
\label{eq:1DHam}
\end{equation}
with proper normalizations, like setting $\hbar=\ell =1$. Eq. (\ref{eq:1DHam}) can be obtained by adding
\begin{equation}
V(x, y) = \frac{y^2}{2} + \tilde{v}(x)
\label{eq:2D Potential}
\end{equation}
to $T$ of Eq. (\ref{eq:hamiltonianintermsofmechanicalmomentum}) and then projection to (say) the LLL, provided
$P_0 \tilde{v}(x) P_0 = v(R_x)$. Note in general $P_0 \tilde{v}(x) P_0 \ne  \tilde{v}(P_0 x P_0 ) = \tilde{v}(R_x)$.
In the following we consider a specific case in which
\begin{equation}
V(x, y) = \frac{x^2 + y^2}{2}.
\label{eq:HO potential}
\end{equation}
In this case we have $\tilde{v} = v$ up to an irrelevant constant.
This corresponds to the harmonic oscillator problem in the 1D QM problem (\ref{eq:Original 1D Ham}) with $v(x) = x^2/2$. We now demonstrate the power of our understanding by obtaining the Hamiltonian eigenstate wave functions in phase space using the symmetry of the corresponding 2D problem with $H= T + V $, {\em without} solving the Schr\"{o}dinger equation. The key to the ``solution" is the observation that the potential (\ref{eq:HO potential}) is rotationally invariant, as a result of which Hamiltonian eigenstates must be angular momentum eigenstates in the symmetric gauge (where rotation symmetry is maintained) as well. However due to the analyticity constraint discussed above, there is only one angular momentum eigenstate for each non-negative integer $m$ with $f(z) = z^m$, and none for negative integers that is normalizable. We have thus found the phase space Hamiltonian eigenstate wave functions to be\cite{Book}
\begin{equation}
\varphi_m(z) = \frac{1}{\sqrt{2\pi 2^{m} m!} } z^m  e^{-\frac{1}{4}|z|^{2}},
\end{equation}
which are the familiar Bargmann functions\cite{Bargmann}. The eigenenergies can be obtained simply by taking the expectation value of $V(x, y)$. We have thus ``solved" the harmonic oscillator problem using the symmetry of the (phase space formulation of this) problem alone!

\section{ Duality, Self-Duality and Quantum Symmetry}

To gain deeper understanding of this symmetry,
we now take a little digression and briefly review the concept of duality in both classical and quantum physics\cite{DualityReview}. A duality transformation refers to a {\em non-local} change of variable that allows for two (usually different) mathematical descriptions of the same physical system or theoretical model. Perhaps the best known example of such duality is the Kramers-Wannier duality of the classical 2D Ising model, and its quantum analog, 1D transverse-field Ising model\cite{DualityReview}.
It plays a crucial role in our understanding of strongly interacting field theory, string theory, and in particular many-body systems near quantum criticality\cite{DualityReview}.

It is perhaps not widely appreciated that ordinary quantum mechanics provides an elementary example of such duality, when one goes from the usual position basis in which the Hamiltonian takes the form
\begin{equation}
H= -\frac{1}{2}\frac{d^2}{dx^2} + v(x),
\label{eq:real space H}
\end{equation}
to the momentum basis in which
\begin{equation}
\tilde{H}= \frac{p^2}{2} + v\left(i\frac{d}{dp}\right).
\label{eq:momentum space H}
\end{equation}
$H$ and $\tilde{H}$ are dual descriptions of the same Hamiltonian but look quite different in general. The reason we normally prefer to work with the real space version (\ref{eq:real space H}) is because it is {\em local}, while for a generic $v(x)$ Eq. (\ref{eq:momentum space H}) represents a highly non-local Hamiltonian in momentum space, which, while still well-defined, is something we prefer to avoid due to our lack of intuition with them. But for special $v$'s like $v(x) = x^n$ for (small) integer values of $n$ this (non-local) change of basis (or Fourier transformation) is a perfectly legitimate duality transformation, yielding an ordinary differential operator just like (\ref{eq:real space H}).

As already mentioned such duality transformations usually yield different looking theories, like Eqs. (\ref{eq:real space H}) and (\ref{eq:momentum space H}) for generic $v$. But there are special cases in which they yield the {\em same} theory (but in general with different parameters). Such self-duality, when present, can be extremely useful; for example the 2D Ising model is self-dual under the Kramers-Wannier duality transformation, and this allowed for the determination of its critical temperature before its exact solution was known, using the fact that the duality transformation must yield {\em exactly the same} Ising model at the critical temperature. As the reader must have realized by now the harmonic oscillator, with $v(x) = x^2/2$, serves as a perfect example of such self-duality as Eqs. (\ref{eq:real space H}) and (\ref{eq:momentum space H}) take exactly the same form in this case, just like the 2D Ising model at its critical temperature! Among other consequences this implies eigen-wave functions must be self-dual under Fourier transformation.

Obviously such self-duality is a symmetry\cite{Zohar}, since the Hamiltonian is invariant under the duality transformation, which is nothing but a change of basis that is realized via a unitary transformation. However this is a somewhat unusual and perhaps unfamiliar type of symmetry, because this change of basis is highly non-local. To illustrate this point let us compare it with more familiar examples of symmetry transformation like translation $x\rightarrow x+l$, or spatial rotation above 1D. In these cases we are actually using the {\em same} real space basis, but only {\em relabelling} the basis states! Obviously there is nothing non-local with such transformations. A slightly less trivial example is spin-rotation, which mixes up states in, say the $S_z$ basis, albeit only locally. But if one uses coherent state path integral, this again reduces to a relabelling of the (coherent) basis states. More formally, in these examples the symmetry transformations leave the classical Lagrangian invariant, and are thus present at the classical level. The self duality symmetry, on the other hand, is invisible (or at least hidden deeply) in the classical Lagrangian. As a result such symmetry was termed quantum symmetry\cite{DualityReview}. We thus found the harmonic oscillator has this quantum symmetry, in addition to the (more obvious) classical symmetries like parity.

Once again harmonic oscillator is very special, even in this regard. Taking the literal meaning of the word duality, one would expect the self-duality symmetry to be a $Z_2$ or Ising type symmetry. However the $x\leftrightarrow p$ duality transformation discussed above turns out to be just a member of a continuous family of such non-local transformations:
\begin{equation}
x\rightarrow x\cos\theta +  p\sin\theta, \hskip 1cm p \rightarrow p\cos\theta -  x\sin\theta,
\label{eq:continuous Duality trans}
\end{equation}
parameterized by the angular parameter $\theta$, and the corresponding basis state wave function written in the original position basis is
\begin{equation}
\phi_{\theta k}(x) = \frac{1}{\sqrt{2\pi}}e^{ik[x - (x^2\cos\theta)/2]/\sin\theta},
\end{equation}
where $k$ is the eigen value of $x\cos\theta +  p\sin\theta$, and the $x\leftrightarrow p$ transformation corresponds to $\theta = \pi/2$. It is obvious that
\begin{equation}
H_{ho} = \frac{p^2 + x^2}{2}
\label{eq:HO H}
\end{equation}
is invariant under the transformation (\ref{eq:continuous Duality trans}). We thus find $H_{ho}$ has a continuous set of self-dualities.

We now return to the discussion of phase space QM. It should be clear by now this continuous self-duality symmetry of $H_{ho}$ is nothing but the rotation symmetry of (\ref{eq:HO potential}). We thus find by formulating QM in phase space with the way we are doing it, this (somewhat abstract) quantum symmetry becomes a very familiar classical symmetry! And this is not too surprising --- in order to formulate QM in phase space, we have to enlarge the Hilbert space by turning the phase space into an enlarged configurational space, which supports more classical symmetry operations. We argue this is an advantage of the phase space formulation of QM.

Before we leave the topic of duality let us consider a less trivial example, in which
\begin{equation}
v(x) = -x^2/2
\label{eq:inverted HO potential}
\end{equation}
known as the inverted harmonic oscillator potential, which is still attracting research activities\cite{InvertedOscillator}. In this case the $x\leftrightarrow p$ duality transformation is almost a symmetry but not quite --- it reverses the sign of the Hamiltonian. It is actually somewhat similar to the particle-hole or chiral symmetry, whose generator anti-commutes (instead of commutes as in ordinary symmetry) with Hamiltonian. As a result it entails certain relations between a state with energy $E$ and its particle-hole partner, whose energy is $-E$, namely their wave functions are related to each other by Fourier transformation. However, for the special case of $E=0$, this becomes an ordinary symmetry, because the Schr\"{o}dinger equation
\begin{equation}
H|\psi\rangle = 0
\end{equation}
is indeed invariant under this duality transformation that results in $H\rightarrow -H$. As a result the usual consequences of symmetry follows, for example the $E=0$ wave functions should be self-dual (or invariant) under Fourier transformation just like harmonic oscillator eigen wave functions. Comparing with the 2D Ising model, here energy plays a role similar to temperature under the duality transformation, and $E=0$ corresponds to the critical temperature which is self-dual.

Now let us inspect this duality transformation and corresponding symmetry consequences in the phase space formulation. This corresponds to
\begin{equation}
V(x, y) =(y^2 -x^2) /2
\label{eq:saddle point potential}
\end{equation}
in (\ref{eq:2D Potential}),
which is a saddle point potential studied by Fertig and
Halperin\cite{FetigHalperin} in the context of integer QHE. This problem is of importance in the understanding of localization
properties of electrons in a magnetic field\cite{HaldaneYang}. Now the $x\leftrightarrow p$ duality transformation corresponds to a transformation $x\leftrightarrow y$ in the 2D plane, which is a standard reflection. Thus the (quantum) self-duality symmetry becomes a (classical) reflection symmetry at $E=0$, which means the phase space wave function should be invariant (up to an overall phase) under such reflection. One immediate consequence is an $E=0$ particle approaching the saddle point has equal probability of turning left and right. This implies in the corresponding 1D problem an $E=0$ particle has half/half probability for transmission and reflection, which is not entirely obvious in the usual real space formulation of QM.

\section{Analyticity of the phase space wave function and its potential utilities}

As is clear from Eq. (\ref{eq:analyticallowest Landau levelwavefunction}), the LLL or equivalently, phase space wave function takes an analytic form. This means $f(z)$ can be expanded as combinations of monomials $z^k$, and the resultant polynomial is uniquely determined by the positions of its nodes, and the number of nodes in $f(z)$ is the same as the degree of the polynomial\cite{Book}. For an infinite 2D plane of the LL problem and corresponding unconstrained phase space for a 1D system, a generic $f(z)$ involves infinite number of monomials and this property is not particularly useful. On the other hand once such spaces are compactified (say into a finite-size sphere or torus), then the number of nodes is fixed to be the number of flux quanta (or LL degeneracy) of the 2D problem, which corresponds to the size of the Hilbert space supported by the phase space of the 1D system. Now the quantum mechanical state of the particle is determined by the positions of these nodes. It is interesting to note that in CM the state is given by the particle's precise position in phase space; in QM this is no longer possible due to the constraint from uncertainty principle, but the precise locations of the nodes, which correspond to positions in phase space where the particle will {\em not} be found, are well-defined, and they determine the state that the particle is in!

We can push this observation further and suggest the following way to formulate QM in a way to make it look as close to CM as possible\footnote{Note one of the motivations to formulate QM in phase space is to make closer contact with CM; see Ref. \onlinecite{Torres-Vega and Frederick}.}. In CM the dynamics of a particle is determined by the (set of two in 1D) Hamilton-Jacobi equations, which are first-order differential equation in time and determine its trajectory in phase space. In QM time-evolution of the state (or wave function) is given by another first-order differential equation in time, the Schr\"{o}dinger equation. Since the phase space wave function is determined by the positions of their nodes, it must be possible to formulate the Schr\"{o}dinger equation as a set of equations that determine the trajectory of the $N$ nodes, where $N$ is the size of the Hilbert space. And that is the reason QM is much harder than CM!

To make the discussion above more concrete let us consider a 1D Bloch electron confined to a specific Bloch band whose dispersion is $\epsilon(k)$, where $-\pi/a \le k < \pi/a$ is restricted to the 1st Brillouin zone, and $a$ is the periodicity of the periodic potential that gives rise to the Bloch bands, or lattice constant. We further confine the electron in real space to $0 \le x < L$, and impose periodic boundary condition, thus making the configuration space a ring. As a result the phase space $(x, k)$ is indeed compact and has the torus topology. The corresponding size of the Hilbert space is $N=L/a$. In the presence of an additional external potential $v(x)$ (which is {\em not} part of the periodic potential and smooth on the scale $a$), the effective Hamiltonian is
\begin{equation}
H = \epsilon(p) + v(x).
\label{eq:band Hamiltonian}
\end{equation}
Formulating this problem in phase space, the state of the electron is described by a LLL wave function of the form (\ref{eq:analyticallowest Landau levelwavefunction}), with additional constraints imposed by the magnetic periodic conditions that correspond to a finite-size system enclosing $N$ magnetic flux quanta\cite{Book}. Such wave functions have $N$ nodes, whose positions [which are actually independent of the gauge choice that led to the form (\ref{eq:analyticallowest Landau levelwavefunction})] determines the state.

Let us first check the consistency of the above by counting the number of independent variables. To determine the positions of $N$ nodes in the 2D phase space we need $2N$ real parameter or equivalently, $N$ complex parameters. Naively this would match the $N$ complex coefficients when expanding a state in an $N$-dimensional Hilbert space. However the overall magnitude and phase of the state are meaningless, thus only $N-1$ complex coefficients are needed. As it turns out the center of mass of the $N$ nodes is actually fixed by the boundary condition angles, as a result the state is already determined by positions of $N-1$ nodes once the boundary conditions have been specified\cite{HaldaneRezayi}. We thus have the correct number of independent variables. At this point we already see parameterizing the wave function in terms of its nodes is more efficient than the wave function itself: Not only is the irrelevant overall magnitude and phase automatically factored out, the node positions are independent of gauge choice, as a result of which the equations below are completely gauge invariant.

In this ``node" representation, the time-dependent Schr\"{o}dinger equation reduces to a set of $N-1$ equations of the form
\begin{equation}
\partial_t\vec{R}_i = \vec{F}_i(\{\vec{R}_j\}),
\label{eq:time-dependent Sch Eq in phase space}
\end{equation}
where $\vec{R}_i$ is the phase space position of the $i$th node, and the ``force" $\vec{F}_i(\{\vec{R}_j\})$ on it depends on the positions of all the nodes and the Hamiltonian (\ref{eq:band Hamiltonian}) in complicated but well-defined ways. We argue Eq. (\ref{eq:time-dependent Sch Eq in phase space}) can be viewed as a quantum analog of the Hamilton-Jacobi equation.
Compared to the Heisenberg equations of motion for observables, here only real c-number variables are involved. Steady states, which correspond to solutions of the time-independent Schr\"{o}dinger equation, satisfy
\begin{equation}
\vec{F}_i(\{\vec{R}_j\}) =0,
\label{eq:time-independent Sch Eq in phase space}
\end{equation}
the solutions of which yield the energy eigenstates, and the eigenenergies can be obtained by taking the Hamiltonian expectation values.

\section{Classical Limit and Quantum Liouville Equation}

The standard way of making connection with CM is to analyse the motion of a wave packet. This might seem hard to do in the present formulation in terms of the node positions, which are where the wave function vanishes instead of peaks. In fact it is very easy to construct a well-localized wave packet by placing the $N-1$ nodes together, say at $(x, k)$, which corresponds to a wave packet localized at its opposite point $([x +L/2], [k+\pi/2])$, where $[\cdots]$ stands for the reduced value after subtracting an integer multiple of the period $L$ and $2\pi/a$ respectively. This is because the nodes act like repulsive charges that push the particle away from them in the so-called plasma analogy of LLL wave functions\cite{Book}. If these nodes were to stick together, the $N-1$ equations of motion (\ref{eq:time-dependent Sch Eq in phase space}) would reduce to one, corresponding to nothing but the classical Hamilton-Jacobi equation once we replace $\vec{R}$ by the location of corresponding wave packet peak. This is how CM emerges in our formulation.

Of course these nodes will spread out under time evolution, which correspond to wave packet dispersion in QM. As a result it is technically hard to keep track of the positions of all the nodes, when $N$ is large. This is what happens in the classical limit of $\hbar\rightarrow 0$, as a result of which the density of states and thus density of nodes in phase space become very large\footnote{In our earlier discussions we have set $\hbar = 1$ and we did not need to distinguish between momentum and wave vector. Now we do.}. In this case instead of tracking the precise locations of these nodes, it is more sensible to coarse-grain the phase space and keep track of the density of nodes, $\rho(\vec{R})$, and the ``Schr\"{o}dinger equation" (\ref{eq:time-dependent Sch Eq in phase space}) reduces to a ``Quantum Liouville equation" of the form
\begin{equation}
\partial_t\rho(\vec{R}) = L[\rho(\vec{R})],
\label{eq:Quantum Liouville Equation}
\end{equation}
where $L$ is the Louville operator. We leave the pursuit of this approach to future work.

\section{Conclusions}

To summarize, we have shown in this paper that lowest Landau level (LLL) projection provides a physical realization of the phase space formulation of quantum mechanics {\em a l\'{a}} Torres-Vega and Frederick\cite{Torres-Vega and Frederick}, and the LLL wave functions widely used in studies of quantum Hall effect are actually phase space wave functions. Analytic properties of such wave functions are of great potential utilities, especially in making connections between quantum and classical mechanics.

\section*{Acknowledgments}
The author thanks Ruojun Wang for assistance. Part of this work was performed at Stanford University during the author's sabbatical leave there, and he thanks Profs. Steve Kivelson, Sri Raghu and especially late Shoucheng Zhang for their invitation and hospitality, as well as Stanford Institute of Theoretical Physics and Gordon and Betty Moore Foundation for support.
This work was supported by the National Science Foundation Grant No. DMR-1932796, and performed at the National High Magnetic Field Laboratory, which is supported by National Science Foundation Cooperative Agreement No. DMR-1644779, and the State of Florida.


\begin{references}
\bibitem{note1} Another (perhaps more popular) approach is based on the Wigner function and various related phase space (quasi) distribution functions, which we will not discuss here. Instead we refer to Refs. \onlinecite{Li} and \onlinecite{Chr}
for discussions on the relations between these approaches.
\end{references}
\end{document}